        \documentclass[runningheads]{svmult}

        \usepackage{makeidx} 
        \usepackage{graphicx} 
   \usepackage{subeqnar} 
      \usepackage{multicol} 
         \usepackage{cropmark} 
    \usepackage{physprbb} 
       \makeindex 

\begin{document}
         \title*{Ultracold atoms and Bose-Einstein condensates in optical
   lattices}
         \toctitle{Ultracold atoms and Bose-Einstein condensates
         \protect\newline in optical lattices}
         %
         %
         \titlerunning{Atoms in optical lattices}
         %
         \author{Oliver Morsch
         \and Ennio Arimondo}
         \authorrunning{Oliver Morsch and Ennio Arimondo}
         %
         %
         \institute{INFM, Dipartimento di Fisica, Universit\`{a} di Pisa, Via
         Buonarroti 2, I-56127 Pisa, Italy}

         \maketitle 

         \begin{abstract}
         For ultracold and Bose-condensed atoms contained in periodic optical
         potential wells the
         quantized nature of their motion is clearly visible. The motion of the
         atomic
         wavepacket can also be accurately controlled. For those systems the
         long-range character of the atomic interaction and of the external
         potential play a key role in the quantum mechanical evolution. The
         basic facets of the experimental and theoretical research for atoms
         within optical lattice structures
         will be reviewed.
         \end{abstract}

         \section{Introduction}
         Crystalline samples of cold atoms, now known as optical
         lattices \index{optical lattices}, were initially
         investigated in the dissipative regime, as a tool to provide
         velocity damping, and hence a reduction of the kinetic energy
         of the atomic samples~(for reviews
         see~\cite{jessen96,grynberg96,meacher98,grynberg01}). In fact
         the sub-Doppler cooling regime relies on the action of laser
         beams on the atomic motion in a standing wave configuration.
         The study of that regime also implied the possibility of
         trapping atoms in the sub-wavelength sized potential wells
         created by the laser beams. As soon it was clear that the
         standing wave patterns created by several intersecting laser
         beams provided by low-power diode lasers could be used to
         trap atoms in periodic structures, the field exploded.  For
         instance a large experimental effort was made to probe
         directly the bound states of the atoms within the optical
         potential.  After the initial experiments using
         one-dimensional (1D) lattices, several schemes were developed
         allowing an extension to two- and three-dimensions (2D and
         3D). Very soon optical lattices were used as a flexible tool
         to modify the spatial periodicity of the cold atomic samples,
         and in some cases fancy spatial structures could be produced
         which the solid state physics community could only dream of.
         Moreover, important applications for atomic
         nanolithography\index{nanolithography} have been realized.

         Later, the research effort moved into the nondissipative, or
         conservative, regime, with the aim of reducing the scattering
         rate in the optical potential wells which ruled out coherent
         control over the wave-packet atomic motion. In fact, the
         interest in optical lattices shifted to using them as a
         test-bed for quantum mechanics. Such a shift in interest was
         enhanced when ultracold atomic samples represented by quantum
         degenerate gases were available for loading into the optical
         periodic potential. Bose-Einstein
         condensates\index{Bose!Einstein condensation} (BEC's)
         represent flexible sources whose spatial dimensions and
         velocity spread can be controlled with large freedom, so that
         a condensate may be loaded with great accuracy into the
         periodic potential created by intersecting laser beams. Thus
         the study of conservative optical lattices used to modify the
         spatial macroscopic wavefunction of BEC's has greatly
         expanded in the last few years.

         In this work we will report on the most important aspects of the
         interaction between ultracold atoms\index{ultracold atoms}, above and below the BEC
         temperature, within optical lattices. We will concentrate on those
         features more directly connected with the long-range interactions
   within the optical
         lattice. We will not, therefore, discuss some exciting investigations
   on the dynamical tunneling and chaotic behavior for atoms located within an
         optical periodic potential whose amplitude is periodically or randomly
         modulated~\cite{steck01,hensinger01}. Furthermore, we will not discuss
         the use of optical lattices in quantum computation
         schemes~\cite{deutsch98}. Moreover the subject of self
         generated periodic spatial structures will not be covered because
         separately treated in this book~\cite{kurizki02}.

         Section~\ref{basicaa} will introduce the basic notions on the creation
   of  optical lattices and the atomic response within the lattice,
         discussing first the near-resonant optical lattices that evolved from
      laser cooling schemes and then the non-dissipative, far-detuned optical
      lattices. Section~\ref{ultracold}
         briefly reviews some of the
         experiments on ultracold atoms in optical lattices carried out since
         1992, whilst section~\ref{condensed} deals with the more recent
      experiments
        in which Bose-Einstein condensates within optical lattices have been
         explored. The experimental results obtained by the Pisa group
         on Bloch oscillations, Landau-Zener tunneling and optical
         potential renormalization are there reported. Section~\ref{conclude}
   concludes the presentation.

         \begin{figure}[b]
         \begin{center}
         \includegraphics[width=.4\textwidth]{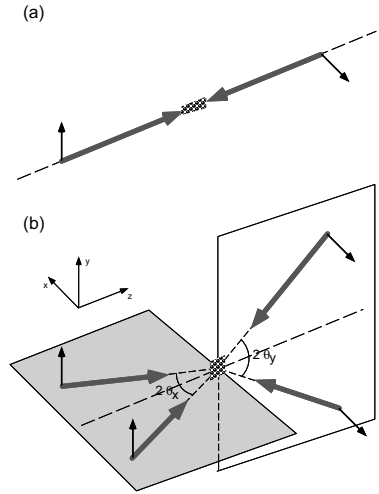}
         \end{center}
         \caption[]{Optical lattices in one and three dimensions. In both
   geometries, a frequency
         difference $\delta$ between the lattice beams can be used to create
          a moving or uniformly accelerated
         lattice.}
         \label{latticegeometries}
         \end{figure}

         \section{Basic notions}\label{basicaa}
      \subsection{Laser cooling}
         The simplest possibility to create a periodic potential for
         neutral atoms is to exploit the light-shift experienced by the
         atoms in a spatially modulated light field. In one dimension, this
         can be achieved by superposing two linearly polarized,
         counter-propagating laser beams with parallel or perpendicular
      polarizations (see
         Fig.~\ref{latticegeometries}(a)).

\begin{figure}[b]
         \begin{center}
         \includegraphics[width=.4\textwidth]{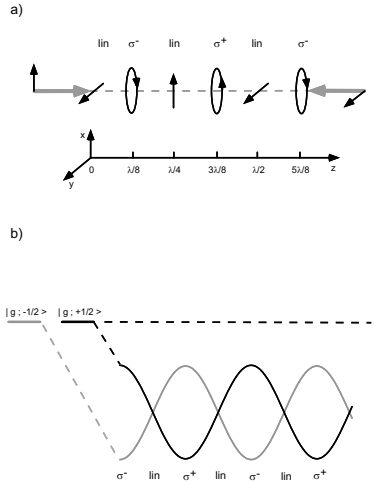}
         \end{center}
         \caption[]{A one-dimensional optical lattice in the $lin\perp lin$
         configuration.} \label{latticeprinciple}
         \end{figure}

\begin{figure}[b]
         \begin{center}
         \includegraphics[width=.4\textwidth]{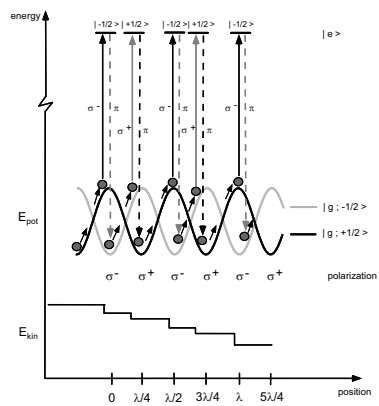}
         \end{center}
         \caption[]{The principle of Sisyphus cooling in a 1D optical
         lattice.} \label{sisyphus}
         \end{figure}

         If the polarizations of the two laser beams are perpendicular
         to each other (see Fig.~\ref{latticeprinciple}), an atom with
         two magnetic sub-levels in the ground state will see two
         interleaved standing waves of $\sigma^+$ and $\sigma^-$ circularly
         polarized light. This so-called $lin\perp lin$ configuration was
         typical of the early experiments on optical lattices, as it
         provided both localization of the atoms at the troughs of the
         potential wells and a sub-Doppler cooling\index{sub-Doppler
           cooling} mechanism ("Sisyphus-cooling", in which atoms are
         preferentially pumped from a sub-level with locally high
         potential energy to the other sub-level with a potential
         minimum at that point, thus reducing the kinetic energy of
         the atoms as shown in Fig.~\ref{sisyphus}.).  For this
         combination of effects to work, the laser beams creating the
         optical lattice were detuned by a few natural linewidths from
         the atomic resonance (near-resonant optical lattices). In
         most configurations, the detuning was to the red side of the
         resonance, resulting in the atoms being trapped at the
         antinodes of the standing wave creating the lattice.

         The 1D periodic structure created by
         two-counter-propagating laser beams can be generalized to two and
         three dimensions in different ways. Theodor H{\"a}nsch and his group
         in Munich used orthogonal pairs of beams whose relative phases
         were stabilized to create a constant lattice
         geometry~\cite{hemmerich93}. A simpler approach was pioneered by
         Gilbert Grynberg's team at the ENS in Paris~\cite{grynberg93}.
         Figure~\ref{latticegeometries}(b) shows their beam geometry used to
      create a 3D lattice. In this setup, no active phase-control
         is necessary as a variation in phase of the beams will only lead
         to a spatial translation of the lattice without changing its
         intrinsic geometry. In 2D and 3D lattices, the
         arrangement of lattice wells can either be anti-ferromagnetic,
         with adjacent wells having orthogonal circular polarizations (as
         in the 1D example shown above), or ferromagnetic, in
         which case adjacent wells have the same circular polarization.

         By changing the splitting angle $\theta$ between the two
         pairs of beams, the distance between neighbouring lattice wells in
         the ENS setup could be varied. This possibility has later also been
   used
      for experiments on BECs. In a 1D optical lattice with angle
         $\theta$, the distance between neighboring wells (lattice
          constant) $d$ can be varied
          through the angle $\theta$ between between the two laser beams
          creating a lattice with
         \begin{equation}
         d=\frac{\pi \sin(\theta/2)}{k_{L}},
         \label{angle}
     \end{equation}
          where $k_{L}$ is the laser wavenumber.

          \subsection{Conservative potential}
          After the exploration of the properties of near-resonant
          lattices, the research effort was concentrated on
          far-detuned lattices with detunings ranging from hundreds to
          thousands of linewidths, using both ultra-cold
          atoms\index{ultracold atoms} and
 BECs\index{Bose!Einstein condensation}. In these optical
 lattices the dissipative
          cooling mechanisms are not active and the optical lattice is
          described by a conservative potential. For a 1D lattice
          configuration as in Fig.\ref{latticegeometries} with the two
          counterpropagating laser beams having the same polarization,
          the ac-Stark shift created by an off-resonant interaction
          between the electric field of the laser and the atomic
          dipole results a potential of the form
            \begin{equation}
            U(x)=U_0\sin^2(\pi x/d),
            \label{potential}
        \end{equation}
         where $U_0$ is the depth of the potential and $d$ the lattice
         constant. In a lattice configuration in which the two laser beams with
         wavevector $k_L$ are counter-propagating, the usual choice of
         units
            are the recoil momentum $p_{rec}=\hbar k_L = Mv_{rec}$ and the
            recoil energy $E_{rec}=\hbar^2 k_L^2/2M$. In the case of an
            angle-geometry, it is more intuitive to
            base the natural units on the lattice spacing $d$ and the
   projection
         $k=\pi/d$ of the laser
         wavevector $k_L$ onto the lattice direction. However, the spatially
            periodic external potential leads naturally to a solid state
            physics approach. One can then define a
         Bloch momentum
         \begin{equation}
         p_B = \frac{2\hbar \pi}{d}=Mv_B,
         \end{equation}
         corresponding to the full extent of the
         first Brillouin zone~\cite{kittel} or, alternatively, to the net
   momentum
         exchange in the lattice direction between the atoms and the
         two laser beams. In that frame of reference a possible choice for the
         energy unit is
         the Bloch energy defined as $E_{\rm B}=\hbar^2 (2\pi)^2/Md^2$.
         These units can also be used for the case of the
         angle-geometry, making use of Eq. (\ref{angle}) for the
         connection between $d$ and $\theta$.

        By introducing a frequency difference $\delta$ between the two
          beams, the lattice potential of Eq. (\ref{potential})
          can be moved at a constant velocity $v_{lat}$
          given by
          \begin{equation}
          v_{lat}=d\delta
          \end{equation}
         or accelerated with an acceleration $a$ given by
         \begin{equation}
         a=d\frac{d\delta}{dt}.
         \label{acceleration}
         \end{equation}

         \section{Ultracold atoms}\label{ultracold}
         \subsection{Early experiments: Local properties}
         The first experiments on optical lattices were aimed at local
         properties of the atoms trapped in the wells. Making a harmonic
         approximation to the potential at the centre of each well, it is
         straightforward to calculate the harmonic trapping frequency for a
         one-dimensional lattice as
         \begin{equation}
         \omega_{harm}=2\frac{E_{rec}}{\hbar}\sqrt{\frac{U_0}{E_{rec}}}.
         \end{equation}

\begin{figure}[b]
         \begin{center}
         \includegraphics[width=.4\textwidth]{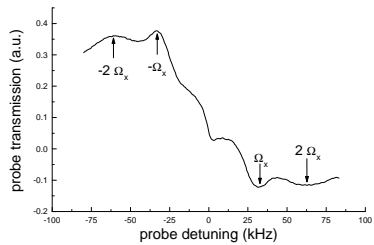}
         \end{center}
         \caption[]{Raman resonances in pump-probe spectroscopy along
         the $x$ axis.}
         \label{raman}
         \end{figure}

         In a typical optical lattice experiment, atoms were first
         trapped and
         laser-cooled in a magneto-optical trap (MOT). After further
         cooling using optical molasses (essentially by switching off the
         magnetic fields of the MOT and increasing the detuning of the trap
         beams), the lattice laser beams were switched on. In order to
         demonstrate that the atoms were truly localized in the potential
         wells of the lattice, a series of experiments was carried out
         using pump-probe techniques~\cite{verkerk92}. By propagating a
         probe beam through the optical lattice that had a variable
         detuning from the lattice beams, Raman resonances could be
         identified (see Fig.~\ref{raman}). These corresponded to an atom
         absorbing a photon from a lattice beam and emitting another into
         the probe beam (or vice versa) whilst changing its vibrational
         quantum number by unity. The measured positions of the Raman
         frequencies agreed with the calculated vibrational frequencies.
         These experiments were helped by the fact that due to Lamb-Dicke
         suppression of inelastic scattering, the Raman resonances were
         narrow enough to be individually resolvable.

         Other experiments investigating the motion of the atoms in the
         lattice wells included the observation of breathing oscillations
         of the atomic wavepackets when then the well-depth was suddenly
         changed~\cite{rudy97} as well as the collapse and revival of
         coherent wavepacket oscillations induced by rapidly shifting the
         lattice in space~\cite{raithel98}. In both experiments, the
         wavepacket motion was inferred from the fluorescence light as the
         re-distribution of photons between the lattice beams due to the
         motion of the atoms led to intensity fluctuations that could be
         experimentally detected. Stimulated revivals were observed in 2000
         by the lattice group in Hannover~\cite{buchkremer00}.

         Information about coherence times of the atomic motion in optical
         lattices can also be obtained by\index{coherent!transients} creating coherent
         transients~\cite{triche96}. In this method, coherences between
         vibrational states are first created by a short probe pulse.
         Subsequently, during the lifetime of the coherence lattice photons
         will be preferentially scattered in the direction of the probe
         beam (now switched off). From the spectrum of these photons,
         vibrational frequencies and coherence times can be
         extracted~\cite{morsch00}.

         \subsection{Global properties}
\begin{figure}[b]
         \begin{center}
         \includegraphics[width=.4\textwidth]{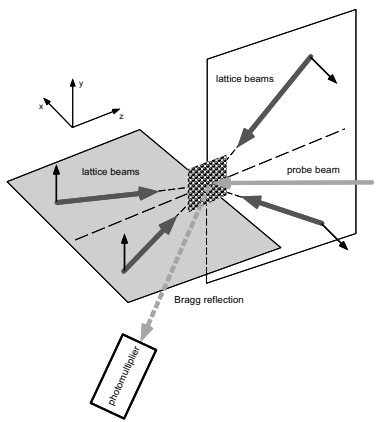}
         \end{center}
         \caption[]{Bragg scattering experiment on an optical lattice. From
         the Bragg-reflected probe beam, information about the spatial
         order and the localization of the atoms can be obtained.}
         \label{bragg}
         \end{figure}

         The nick-name "crystals bound by light" that was invented for
         optical lattices soon after their first experimental realization
         highlighted a property of these physical systems that was not
         visible in the early experiments, namely their periodic spatial
         structure. If optical lattices were, as predicted, "egg cartons"
         for atoms, then some direct evidence to that effect was desirable.
         An obvious experiment to carry out was Bragg-scattering\index{Bragg scattering} of a probe
         beam off the planes of the "crystal", and in 1995 the groups in
         Munich and Gaithersburg~\cite{weidemuller95,birkl95} managed to do
         just that (see Fig.~\ref{bragg}). They were able to show that when the
         lattice beams were switched on, long-range spatial order was built
         up as the atoms were further cooled and localized in the
         periodically arranged potential wells of the lattice. When the
         lattice was switched off, the spatial order was lost on a
         time-scale consistent with the thermal motion of the atoms. By
         measuring the structure factor associated with the Bragg
         reflection, it was also possible to obtain information about the
         spatial localization of the atoms in the lattice
         wells~\cite{gorlitz97}.

         Another global property of near-resonant optical lattices that
         was
         accessible to experimental verification was their response to external
      magnetic
         fields. Using an optical lattice with anti-ferromagnetic order
         of the atomic spins, the ENS group showed that by applying a
         magnetic field that differentially shifted the potential energies
         of the $\sigma^+$ and $\sigma^-$ wells, a redistribution of
         population between the two sets of wells resulted in a macroscopic
         magnetization of the lattice~\cite{meacher95}. The entire lattice,
         therefore, exhibited paramagnetic behaviour.

         In near-resonant optical lattices, only a few percent of all the
         lattice sites actually contain an atom. Despite the consequent
         lack of atom-atom interactions, it was shown in 1996 that
         propagating excitations similar to sound waves can be created in
         optical lattices~\cite{courtois96}. Through cycles of
         half-oscillations in the potential wells and subsequent optical
         pumping into the other magnetic sub-level of the ground state,
         propagating density modulations were created that could be
         detected via pump-probe spectroscopy. Moreover the first direct
         observation of Brillouin-like propagation modes in a dissipative
   periodic
         optical lattice was recently reported~\cite{sanchez02}.

         \subsection{Quasi-periodic lattices}
         Whilst so far we have stressed the regular, periodic structure of
         optical lattices, a number of interesting experiments have also
         been carried out in a quasi-periodic geometry. By adding an
         additional laser beam to the four-beam geometry described above, a
         quasi-periodic lattice structure can be created. The diffusion
         properties in such a lattice were studied by the group of Philippe
         Verkerk~\cite{guidoni99}.

         \subsection{Far-detuned lattices}
         The fact that in a red-detuned, near resonant lattice the atoms
         are trapped at local maxima of the light intensity means that the
         coherent motion of the atomic wavepackets is frequently
         interrupted by spontaneous emission events. Early on in the
         development of optical lattices, schemes were developed that would
         provide trapping in locally "gray" or "dark" states in which the
         atoms absorb very few photons or no photons at all. These schemes
         were, appropriately, called "gray" or "dark" optical
         lattices~\cite{hemmerich95}.

         In order to get rid completely of dissipative effects, it was,
         however, necessary to increase the detuning of the lattice beams
         from the atomic resonance and, at the same time, the beam
         intensity in order to keep a fixed lattice depth. As the
         spontaneous scattering rate decreases more rapidly with detuning
         than the light-shift or dipole force providing localization, in
         this manner optical lattices could be realized in which for the
         duration of the experiment (usually on the order of tens to
         hundreds of milliseconds) virtually no photons were scattered. The
         lack of photon scattering, on the other hand, also ruled out
         the Sisyphus cooling mechanism present in near-resonant lattices.
         In order further to cool atoms transferred into a far-resonant
         lattice, resolved-sideband Raman cooling\index{Raman cooling} was successfully employed
         by several groups~\cite{hamann98,perrin98,vuletic98}. As in
         trapped ion experiments, transitions towards lower vibrational
         levels were induced using a combination of laser beams and
         magnetic fields. In this way, a large fraction of the atoms could
         be cooled to the ground vibrational state of the lattice.

         Another possibility to achieve the loading of atoms into the
         ground vibrational state (or, for shallow lattices, the ground
         state Bloch band) is to start from a Bose-condensed cloud of
         atoms, as will be discussed in the second part of this chapter.

         \subsection{A Lego-kit for quantum systems} Apart from the intrinsic
         interest in the characteristics
         of optical lattices, a number of experiments have used optical
         lattices as a tool for creating a specific quantum system whose
         properties could then be studied. Especially the precise
         experimental control of the lattice parameters (well depth,
         geometry) means that optical lattices can be used as a
         construction kit for quantum systems.

         In a beautiful experiment~\cite{bendahan96,peik}, the group of
         Christophe Salomon at the ENS in Paris demonstrated Bloch
         oscillations\index{Bloch oscillations} of atomic matter waves delocalized over many lattice
         sites of a shallow optical lattice by accelerating the lattice and
         measuring the resulting velocity of the atoms. Related experiments
         by Mark Raizen's team in Austin investigated\index{Landau!Zener tunneling} Landau-Zener
         tunneling and Wannier-Stark ladders~\cite{madison99}. In an
         extension to this work, they were able to experimentally verify
         the non-exponential nature of the initial decay of a quantum
         system~\cite{wilkinson97} and the associated Zeno and anti-Zeno
         effects exhibited by such systems when they are subject to
         frequent observations~\cite{fischer01}.

         In a near-resonant lattice, Gilbert Grynberg's group at the ENS
         was able to create an asymmetric optical lattice with a
         ratchet-like potential that converted the random thermal motion of
         the atoms into directed motion~\cite{mennerat99}.

         \section{Bose-condensed atoms}\label{condensed}
         While in most of the original optical lattice experiments the atomic
      clouds  had temperatures in the
         the micro-Kelvin range, corresponding to a few recoil energies of the
      atoms, atomic samples with
         sub-recoil energies are now routinely produced in \index{Bose!Einstein condensation}Bose-Einstein
      condensation
         experiments. Since the
         first experimental realizations in 1995, many aspects of Bose-Einstein
         condensed atomic clouds
         (BECs) have been studied~\cite{inguscio}, ranging from collective
         excitations to superfluid
         properties\index{superfluidity} and quantized vortices\index{quantized vortices}.
         The properties of BECs in periodic potentials constitute a vast new
   field of research initially explored in~\cite{sorensen98,jaksch98}.
         Several experiments have made use of the periodic optical potential
         produced by a pulsed standing wave to manipulate the condensate or to
         explore its properties
   ~\cite{kozuma99,ovchinnikov99,hagley99,stenger99,simsarian00,vogels02,steinhauer02,ozeri02}.
         In the following, we will concentrate on studies of the condensate
   within
      the periodic optical
         lattice. The first step in that direction was taken by the
         investigation of the
         tunneling of BECs out of the one-dimensional potential wells of a
   shallow
         optical lattice in the presence of
         gravity~\cite{anderson98}. More
         recently the phase
            properties of the condensate wavefunction occupying the whole
            optical lattice have involved such intriguing concepts as number
            squeezing~\cite{orzel01} and the Mott
            insulator\index{Mott insulator}
         transition~\cite{greiner02}. The tunneling of the condensate
         between neighboring wells, controlled by varying the optical lattice
         potential depth, determine the overall properties of the macroscopic
         wavefunction. Thus the condensate response within a 1D optical lattice
         can be described as an array of tunneling junctions, as pointed out
      in~\cite{anderson98} and later explored
      in~\cite{burger01,cataliotti01}
         in connection with the superfluid\index{superfluidity} properties of the
         condensate wavefunction. Coherent acceleration of BECs adiabatically
      loaded into optical
         lattices was demonstrated in~\cite{morsch01,cristiani02}, with Bloch
         oscillations\index{Bloch oscillations} observed for small values of the lattice depth, while a
         Landau-Zener\index{Landau!Zener tunneling} breakdown occurred when the lattice depth was further
      reduced and/or the acceleration
         increased. The expansion of the condensate array was
         explored initially in~\cite{pedri01} and later in~\cite{morsch02}.
         The high level coherent control over this artificial solid state
         system was demonstrated in~\cite{denschlag02}, where the BEC
         was carefully loaded into the lattice ground state by
         adiabatically turning on the optical lattice. The different dependence
   of the condensate population on the
         temperature for the 1D optical lattice was pointed out by
         ref.~\cite{burger01b}.

         Most experiments were based on the production of the condensate
         through the standard technique, followed by an adiabatic load into the
         optical lattice. Often the magnetic trap was switched off when the
      optical
         lattice was switched on. However, a larger condensate density is
   realized
         when the interaction between the
         condensate and the lattice takes place inside the magnetic trap,
         and both of them are subsequently switched off
         to allow time-of-flight imaging. The Florence group~\cite{burger01}
   has
         developed a different approach by producing the condensate directly
         inside the optical lattice which is adiabatically loaded during the
         evaporating cooling stage. The main advantage of this approach is
         that the phase coherence of the condensate over the whole optical
         lattice structure is built up during the condensate formation
         process. The main disadvantage is that the optical lattice is on
   during
         the whole evaporation process, and the spontaneous losses produced by
         the optical lattice should be heavily reduced by further increasing
   the
      detuning
         of the optical lattice lasers. While a large majority of the
         experiments have loaded the condensate into 1D optical lattices,
         experiments on 2D lattices were performed by Greiner {\it et al.}
         \cite{greiner01}, and the Mott-Hubbard
         transition of ref.~\cite{greiner02} was realized in a
         3D optical lattice\index{Mott insulator}.

          If the momentum spread of the atoms loaded into an optical lattice
         structure
            is small compared to the characteristic lattice momentum $p_{\rm
   B}$,
         then
            their thermal de Broglie wavelength will be large compared to the
            lattice spacing $d$ and will, therefore, extend over many
            lattice sites. A description in terms of a coherent delocalized
         wavepacket within a
            periodic structure is then appropriate and leads us directly to
            the Bloch formalism first developed in condensed matter physics.
        In the tight-binding limit ($U_0\gg10\,E_{\rm
         rec}$), the
            condensate in the lattice can be approximated by wavepackets
            localized at the individual lattice sites (Wannier states). This
         description is more intuitive than
            the Bloch picture in the case of experiments in which the
         condensate is released from a (deep) optical
            lattice into which it has previously been loaded adiabatically.

         \subsection{BEC theory in optical lattices}
         In Bose-Einstein condensates, interactions between the
         constituent atoms are responsible for the non-linear behavior
         of the BEC and can lead to interesting phenomena such as
         solitons~\cite{burger99} and four-wave mixing with matter
         waves~\cite{deng99}. As the atoms are extremely cold, collisions
         between them can be treated by considering only
         $s$-wave scattering, which is described by the scattering length
         $a_s$. Modeling the interatomic interaction as hard-core
         collisions, one can simplify the treatment using a mean-field
         description which leads to the famous Gross-Pitaevskii
         equation~\cite{dalfovo99}, the validity of which has been demonstrated
         in numerous experiments. For a BEC in an optical lattice, one expects
   an
         effect due to
         the mean-field interaction\index{mean field!interaction} similar to the one responsible for
         determining the shape of a condensate in the Thomas-Fermi limit:
         The interplay between the confining potential and the
         density-dependent mean-field energy leads to a modified ground
         state that reflects the strength of the mean-field interaction.
         Applied to a BEC in a periodic potential, one expects the
         density modulation imposed on the condensate by the potential
         (higher density in potential troughs, lower density where the
         potential energy is high) to be modified in the presence of
         mean-field interactions. In particular, the tendency of the
         periodic potential to create a locally higher density where the
         potential energy of the lattice is low will be counteracted by
         the (repulsive) interaction energy that rises as the local
         density increases.

         The description of a condensate in a 1D array of coupled
            potentials wells is based on the total Hamiltonian
            \begin{equation}
         H_{tot}=\frac{-\hbar^2}{2M}\nabla^2+U_0\sin^2(\pi
         \frac{x}{d})+g|\Psi(\vec{r})|^2,
            \label{hamiltonianaa}
         \end{equation}
         with the interaction parameter $g$ given by
         \begin{equation}
         g=\frac{4\pi \hbar^2 a_s}{M},
         \label{g}
         \end{equation}
         and $\Psi(\vec{r})$ the condensate wavefunction at position $\vec{r}$.

          As the interaction term is expected to distort the band structure
          of the condensate in the lattice~\cite{sorensen98}, it should affect
          all measurable quantities (Rabi frequency, amplitude of Bloch
      oscillations,
          and tunneling probability). In Ref.~\cite{choi99}, the authors
   derived
      an
         analytical expression
         in the perturbative limit (assuming $U_0\ll E_B$) for the
         effect of the mean-field interaction on the ground state of the
         condensate in the lattice. Starting from the Gross-Pitaevskii equation
         for the condensate wavefunction in a one-dimensional
         optical lattice ({\it i.e.} a one dimensional Hamiltonian
         equivalent to that of Eq. (\ref{hamiltonianaa})), they found that
         the effect of the mean-field interaction\index{mean field!interaction} could be approximately
         accounted for by substituting the
         potential depth $U_0$ with an effective potential
            \begin{equation}
            U_{eff}=\frac{U_0}{1+4C},
         \label{potentialred}
         \end{equation}
         with the dimensionless parameter $C$ given by~\cite{choi99}
         \begin{equation}
         C=\frac{\pi n_0 a_s}{k_L^2\sin^2(\theta/2)}=\frac{n_{0}g}{E_{\rm B}},
         \label{cparameter}
         \end{equation}
          corresponding to the ratio of
         the nonlinear interaction term and the Bloch energy. The $C$ parameter
         contains the peak condensate density $n_0$, the scattering length
      $a_{s}$\index{s-wave scattering length},
         and the atomic mass $M$. From the dependence of $C$ on the lattice
   angle
         $\theta$ it follows that a small angle $\theta$ (meaning a large
   lattice
         constant $d$) should result in a large interaction term $C$.
         The reduction of the effective potential given by
            Eq. (\ref{potentialred}) agrees
            with the intuitive picture of the back-action on the periodic
         potential of the density
            modulation of the condensate imposed on it by the lattice
            potential. For repulsive interactions, this results in the
            effective potential being lowered with respect to the actual
            optical potential created by the lattice beams.

         \subsection{Theoretical advances}
         The properties of Bose-Einstein condensates within optical lattices
         have been examined in a large number of theoretical papers, predicting
         a variety of phenomena, often making use of the strict analogies with
         cases previously studied within the context of solid state physics and
         nonlinear
         dynamics. We will list here the principal research lines that have
         characterized this research so far.

         The collective excitations of the condensate within optical lattices,
         and their probe, have been
         determined~\cite{brunello00,chiofalo01,kramer02}, and those
         analyses have stimulated the experimental investigations in
         refs.~\cite{vogels02,steinhauer02,ozeri02}. A strong deformation of
         the Bloch energy bands of the condensate produced by the nonlinear
         atomic interactions has been predicted in
         refs.~\cite{bronski01,wu02,diakonov01}. Different mechanisms of
   breaking
      down
         the Bloch
         oscillation, all of them connected to the interatomic interactions,
         have been discussed in refs.~\cite{wu00,zobay00,trombettoni01,wu01}.

         The thermal and quantum decoherence
         for an array of multiple condensates within an optical lattice,
         introduced in refs.~\cite{pitaevskii01,cuccoli01}, are an important
   issue
         requiring more detailed studies, both theoretically and
   experimentally.
         The Bose-Hubbard-Hamiltonian for atoms in an optical lattice,
         introduced by Jaksch {\it et al.}~\cite{jaksch98}, was analyzed in
         ref.~\cite{vanoosten} through
         a mean-field approximation generalization of the Bogoliubov approach;
      later
         it was applied to determine the conditions for the
         number squeezing in that transition~\cite{burnett02}.

         A number of papers have pointed out the existence of additional
         solutions for the evolution of the condensates within optical
   lattices:
         instabilities, solitons (shape preserving
         excitations), breathers (excitations characterized by internal
         oscillations), and self-trapping states or intrinsic localized modes
      (wavepacket
         localized around few
         lattice sites)~\cite{trombettoni01,abdullaev01,konotop01,tsukada02}.
      Those
         predictions have often made use of
         theoretical analogies with
         other nonlinear classical and quantum problems, involving the
   sine-Gordon
         equation, the discrete nonlinear Schr{\"o}dinger equation and other
         nonlinear physics problems. Spatial instabilities
         of the condensate within the optical lattice, with a spontaneous
         breaking in the spatial periodicity, have been predicted by Wu and
         Niu~\cite{wu01}.

         \begin{figure}[b]
         \begin{center}
         \includegraphics[width=.4\textwidth]{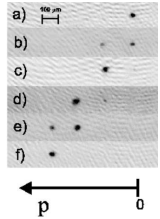}
         \end{center}
         \caption[]{Coherent acceleration of a Bose-Einstein condensate. In
      (a)-(f)
         $U_0=2.3\,\mathrm{E_{\rm rec}}$
         $a=9.81\,\mathrm{m\,s^{-2}}$, the condensate being accelerated for
         $0.1,0.6,1.1,2.1,3.0$
         and $3.9\,\mathrm{ms}$, respectively. The separations between the
      different
         spots
         vary because the detection occurred after different
         time delays.}
         \label{leshouches2}
         \end{figure}

         \subsection{Experimental results}
         \subsubsection{Bloch oscillations}
         For 1D optical lattice, a linear increase of the detuning $\delta$
         between the two laser beams forming the lattice provides to the
   optical
         lattice the constant acceleration given by Eq. (\ref{acceleration}).
   As
      the
         lattice
         can only transfer momentum to the condensate in units of $2\hbar k$,
         the acceleration of the condensate leads to
         higher momentum classes as the acceleration time increases.
         With an initial momentum spread of the condensate much less than
         the Bloch momentum $p_{\rm B}$) and since the adiabatic switching
         transfers the momentum spread into lattice quasimomentum, the
   different
         momentum classes
         $p=\pm np_{\rm B}$ (where $n=0,1,2,...$) occupied by the
         condensate wavefunction can be resolved directly after the
      time-of-flight.
         In the experiments of refs.~\cite{morsch01,cristiani02} with the
         optical lattice in the horizontal direction and the atomic wave
         diffraction monitored in the time-of-flight detection, the accelerated
         momentum
         classes showed up as diffraction peaks in the time-of flight
         absorption images as in Fig.\ref{leshouches2}. Up to $6\,p_{\rm B}$
         momentum could be transferred to the condensate
         preserving the phase-space density of the condensate during the
         acceleration,
         a result indicating that no heating or reduction of the
         condensate fraction
         occurred.
         Measuring the average velocity of the condensate from the occupations
   of
      the
         different
         momentum states, the Bloch oscillations\index{Bloch oscillations} of the condensate velocity
         corresponding to a Bloch-period $\tau_{\rm B}=h/(Ma_{s}d)$ could be
         detected. Note in Fig.\ref{leshouches2} that while the Bloch
         oscillation takes place, the condensate wavefunction coherently
   occupies
         two neighboring velocity classes. In a
         configuration with the optical lattice oriented along the vertical
         direction~\cite{anderson98,cristiani02}, the different momentum
   classes
         emitted from the condensate
         travel in space separately because of the acceleration due to gravity.
      Thus
         the condensate absorption images after the time of flight show
         one or two atom laser pulses corresponding to the single or double
      momentum class
         occupied by the condensate at the time of the release.

         \subsubsection{Landau-Zener tunneling}
\begin{figure}[b]
         \begin{center}
         \includegraphics[width=.4\textwidth]{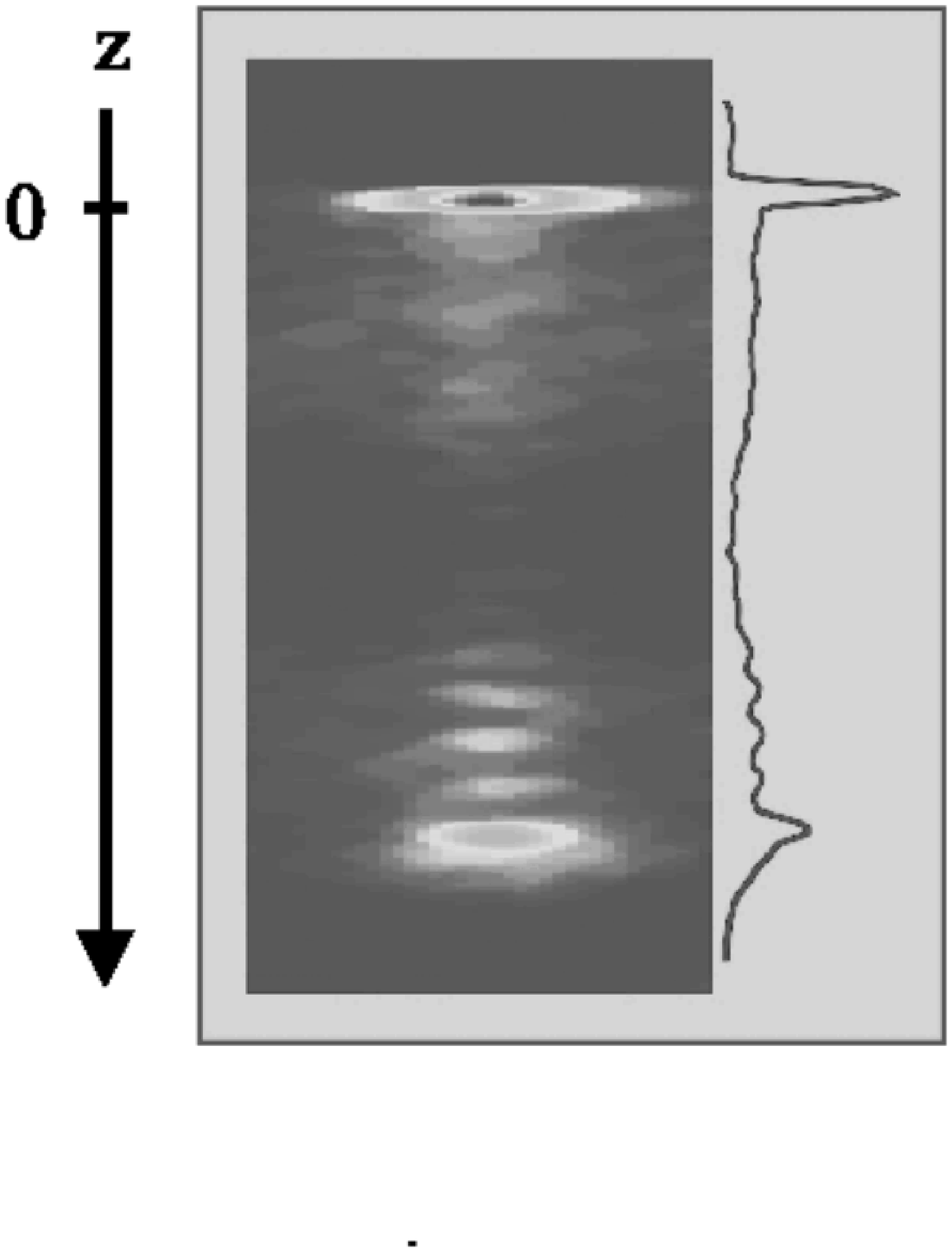}
         \end{center}
         \caption[]{Time-of-flight image and, on the left, transverse
         vertical profile for the Landau-Zener tunnelling of a condensate from
   a
      1-D
         optical lattice oriented vertically. The $p= + |n|p_{\rm
         B}$ atomic momentum classes
         are separated by the gravity, pointing downwards. The large condensate
         amplitude near
         $z=0$ corresponds to the population confined in the lattice. The
         atomic momentum classes generate the atom laser pulses detected after
         a $10.1\,\mathrm{ms}$ time of flight, with
         $U_0=10\,\mathrm{E_{\rm rec}}$ and condensate acceleration for
         $10\,\mathrm{ms}$.}
         \label{atomlaserpulses}
         \end{figure}

         At large acceleration of the lattice or at a decreased lattice depth,
         not all of the condensate could be coherently accelerated up to the
         final velocity of the lattice. Such condensate loss can be interpreted
         in terms of Landau-Zener tunneling\index{Landau!Zener tunneling} of the condensate out of the lowest
      band
         when the edge of the Brillouin zone is reached. Each time the
         condensate is accelerated across this edge, the fraction
         undergoing
         tunneling into the first excited band is given by the Landau-Zener
         probability $P_{\rm LZ}$~\cite{bendahan96,morsch01,denschlag02}:
         \begin{equation}
         P_{\rm LZ}=e^{-a_c/a}
         \end{equation}
         with the critical acceleration $a_c$ given by
         \begin{equation}
         a_{c}=\frac{\pi U_0^2}{16\hbar^2 k}.
         \end{equation}
         Such tunneling produces the following mean velocity $v_{m}$
         of the condensate at the end of the acceleration process for a final
         velocity $v_{\rm B}$ of the lattice:
         \begin{equation}
         v_m=(1-P_{\rm LZ})v_{\rm B}.
         \end{equation}
         In ref.~\cite{morsch01} it was verified that this formula correctly
         described
         the tunneling of the condensate at low values of the condensate
   density
      by
         varying
         both the potential depth and the lattice acceleration. In the
         configuration with the optical lattice oriented along the vertical
         direction~\cite{anderson98,cristiani02}, the different momentum
   classes
         emitted
         from the condensate travel in space separately because of the
      acceleration due to gravity.
         Thus the condensate absorption images after the time of flight
         (Fig.\ref{atomlaserpulses}) showed a part of the
         population confined in the optical lattice, and several
         laser pulses corresponding to the different momentum class
         occupied by the condensate.

         \subsubsection{Optical potential renormalization}
\begin{figure}[b]
         \begin{center}
         \includegraphics[width=.4\textwidth]{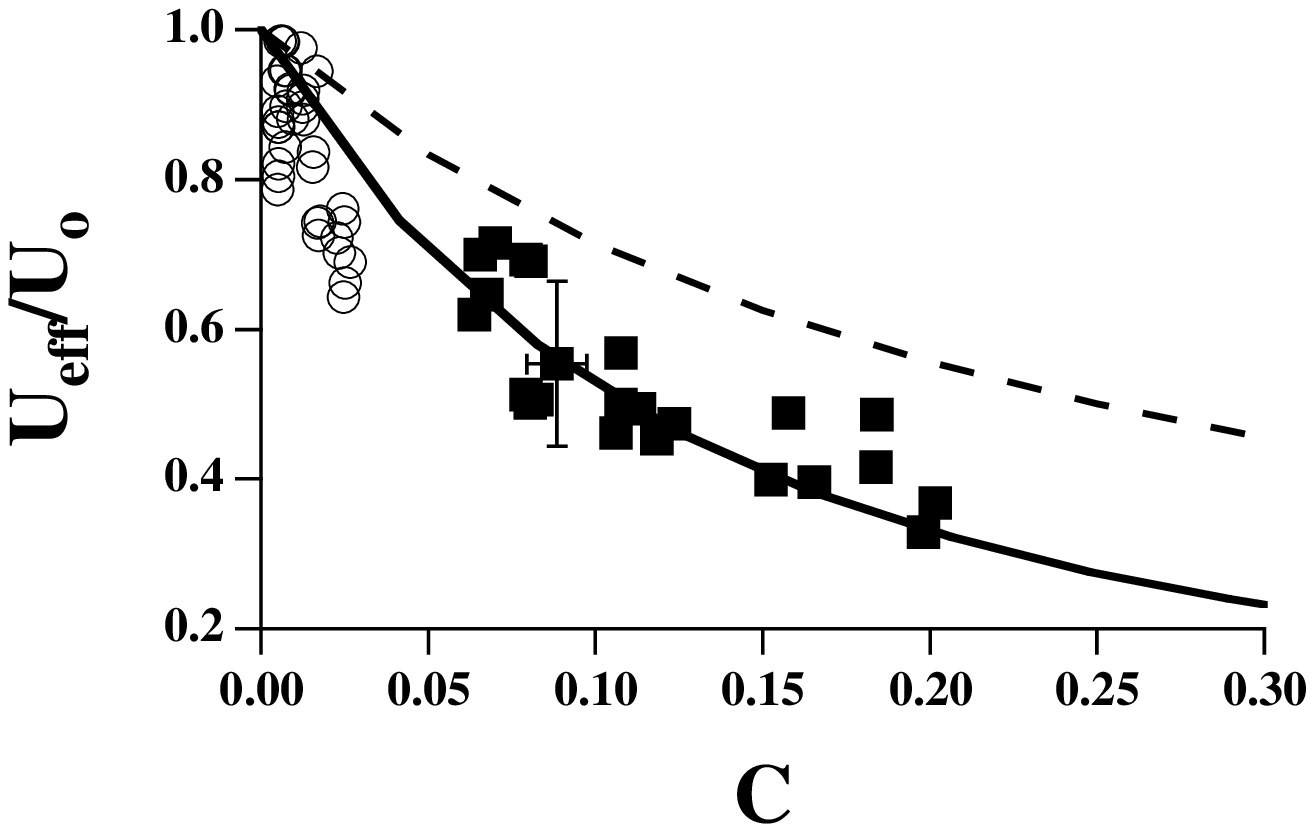}
         \end{center}
         \caption[]{Dependence on the $C$ parameter for the effective potential
         $U_{\rm eff}$, normalized to the applied optical potential
         $U_{0}$,
         for the two lattice geometries of ref.~\cite{morsch01}. The
   experimental
         results for the counter-propagating geomotry
            (open circles) and angle geometry with $\theta=29$ deg (filled
      squares) are plotted together
         with the
         theoretical prediction of ref.~\cite{choi99}(dashed line) and of
         ref.~\cite{kalinski}(continuous line). Parameters in these
         experiments were $a=23.4\,\mathrm{m\,s^{-2}}$ and
         $U_0=2.2\,E_{\rm rec}$ for the counter-propagating lattice and
         $a=3.23\,\mathrm{m\,s^{-2}}$ and $U_0=5.7\,E_{\rm rec}$ for
         the angle geometry.}
         \label{meanfield}
         \end{figure}

         In order to measure accurately the variation of the
         effective potential $U_{\rm eff}$ with the interaction parameter
         $C$ as given by Eq. (\ref{potentialred}), the Landau-Zener
         tunneling\index{Landau!Zener tunneling}
         out of the lowest Bloch band for small lattice depth was
         studied in a regime where the parameter $C$ modified the optical
      potential
         experienced by the condensate~\cite{morsch01,cristiani02}.
         Therefore, the variation of the final mean velocity $v_m$ was studied
         as a function of the condensate density.
         The density was varied by changing the mean harmonic frequency
         of the magnetic trap. From the mean velocity the effective potential
   was
         then calculated using the Landau-Zener probability given above, with
   the
         critical acceleration determined by the effective potential
         \begin{equation}
         a_{c}=\frac{\pi U_{\rm eff}^2}{16\hbar^2 k}.
         \end{equation}
         Fig.~\ref{meanfield} shows the ratio $U_{\rm eff}/U_0$ as a function
   of
         the parameter $C$ for the experimental investigation in two different
         geometries of the optical lattice, counter-propagating and in an
      angle-geometry with $\theta=29$ deg.
         As predicted by Eq. (\ref{cparameter}), the reduction of the effective
         potential is much larger in the angle geometry. The theoretical
      predictions
         of Eq.~(\ref{potentialred}) given by the theory of ~\cite{choi99} are
      also
         shown in
         the figure. An effective potential may also be derived within the
      framework of
         a
         tight-binding approximation using Wannier states, i.e. describing the
         condensate within
         each potential minimum potential through the solution of the
         Gross-Pitaevskii equation while neglecting the overlap with
   neighboring
         potential minima \cite{kalinski}, a very weak tunneling of the
         condensate preserving the overall phase
         relation of the condensate wavefunction. The tight-binding predictions
         presented
         in Fig. \ref{meanfield} provide a better agreement with
         experimental results than those of Eq.(\ref{potentialred}).

         \subsubsection{Squeezed states and Mott insulator}
         Because the condensate is a superfluid, described by a wavefunction
         exhibiting long-range phase coherence, if the lattice potential is
         turned on smoothly, the condensate remains in the superfluid phase.
         In this regime a delocalized condensate wavefunction minimizes the
         total energy of the many-body system, the phase of the atomic
         wavefunction being well determined, with a fluctuating atom number at
         each lattice site.
         This applies as long as the atom-atom interactions are small compared
         to the tunnel coupling. Instead, when the repulsive atom-atom
         interactions are large compared to the tunnel coupling, the total
         energy is minimized when each site of the optical lattice is filled
         with the same number of atoms. Approaching this Mott-isulator quantum
      phase\index{Mott insulator}
         transition, with the lattice site number commensurate to the atom
         number, the wavefunction phase coherence is lost. Meanwhile,
          the fluctuations in the number of atoms per site are
         reduced and finally go
         to zero. In addition, in the
         superfluid\index{superfluidity}
         regime the excitation spectrum is gapless, whereas the Mott insulator
         phase exhibits a gap in the excitation spectrum.

         The first important step in modifying a condensate from a coherent
   state
      to a Fock state was taken by Orzel {\it et al.}~\cite{orzel01} who
   realized
      squeezed states in a 1D optical lattice. By properly choosing the depth
   of
      the optical potential and the amplitude of the mean field interaction
   term
      in Eq. (\ref{hamiltonianaa}), the authors were able to reach a regime
   where
         deviation from the coherent state wavefunction for the condensate
         became significant. An increase in the phase fluctuations of
         the condensate wavefunction was measured in the interference of atom
   waves
         released from the
         optical lattice. From this, a squeezing in the number of atoms
   occupying
         the lattice sites was derived by assuming a minimum uncertainty
         state.

         The experimental realization of the Mott isulator\index{Mott insulator} in a 3D optical
      lattice
         in
         ref.~\cite{greiner02} required a cubic optical lattice where a volume
         with 65 sites in each single direction was occupied, with potential
         depth around $U_{0}=12\,E_{\rm rec}$. Adiabatic loading of a
   condensate of
         $2\times 10^{5}$
         atoms into such a lattice produced a state whose coherence
         properties were tested as usual by the interference pattern following
         a time-of-flight. At a potential depth larger than $10\,E_{\rm rec}$
   the
         interference pattern presented an incoherent background, gaining more
         strength increasing the potential depth until that background was the
         only visible feature in the interference. That loss of interference
   was
         a first sign of the Mott-isulator regime, with a Fock state without
         macroscopic phase coherence characterizing the occupation of the
         single wells. Evidence of the phase transition was gained by
         monitoring the quick restoration of the phase coherence, quick as
      compared
         to the restoration produced by the inhomogeneous dephasing of the
         superfluid\index{superfluidity} condensate wavefunction under the application of a magnetic
         energy gradient. Additional strong evidence was provided by the
         direct measurements of the gap in the excitation spectrum of the Mott
         insulator\index{Mott insulator}, in contrast to the superfluid phase where the excitation
      spectrum
         is gapless.

         \section{Conclusion and Outlook}\label{conclude}
         We have provided a brief overview of the research work performed on
   cold
         atomic samples (laser cooled or evaporatively cooled to the quantum
         degenerate regime) located within the periodic potential created by an
         optical interaction, concentrating our attention on the phenomena
         where the long range order of the external potential is imprinted
         onto the atomic wavefunction. As far as
         quantum degenerate gases within optical lattices are concerned,
   recently
      the
         field has greatly flourished with new experimental groups joining the
         crowded space and theory groups proposing new exciting schemes and
         pointing out the occurrence of interesting phenomena. It seems
         that
         Bose-Einstein condensates in optical lattices could be the
         test-bed
         for a large variety of theoretical models.
          It is obvious that this flourishing of ideas and of measurements will
          continue still for
         some time, because additional configurations could be explored, the
   control on the atoms and the optical lattices will progress, and
         finally because other phenomena may occur when Fermi
         degenerate gases will be loaded in optical lattices. The detailed
         comparison between theory and experiments dealing with Bose-Einstein
         condensates within optical lattices requires heavy numerical
         solutions of the 3D Gross-Pitaevskii equation. However, the recent
         realization of nearly 1D Bose-Einstein condensates could
         simplify the theoretical analysis.

         But even with non-condensed, ultra-cold gases the versatility of
   optical
      lattices in realizing and studying
         quantum systems will certainly provide interesting challenges to
         physicists in the future. Studies of cold collisions and diffusion
         properties in optical lattices are but two of the areas in which a
         lot of work still needs to be done. Moreover, experiments will test
         the schemes in which the coherent motion of the atoms in far-detuned
         optical lattices could be exploited to realize quantum logic gates.

         \section*{Acknowledgments}
         The collaboration of D. Ciampini, M. Cristiani and J.H. M{\"u}ller to
         the experimental results reported in the present work is gratefully
         acknowledged. This work was supported by the MURST through the
         PRIN2000
           Initiative, by the INFM through the Progetto di Ricerca Avanzata
           `Photon matter', and by the the EU through the Cold Quantum
           Gases Network, Contract No. HPRN-CT-2000-00125. O.M. gratefully
           acknowledges a Marie Curie Fellowship from the EU within the IHP
           Programme.

         %


\begin{thebibliography}{8.}
         \addcontentsline{toc}{section}{References}
         \bibitem{jessen96} P.S.~Jessen and I.H.~Deutsch, Adv. At. Mol. Opt.
   Phys.
         \textbf{37}, 95 (1996).
         \bibitem{grynberg96} G.~Grynberg and C.~Trich{\'e}, in {\it Coherent and
         Collective
         Interactions of Particles and Radiation Beams} edited by A. Aspect,
         W. Barletta, and R. Bonifacio (IOS Press, Amsterdam) (1996).
         \bibitem{meacher98} D.R.~Meacher, Cont. Phys. \textbf{ {39}}, 329
         (1998).
         \bibitem{grynberg01} G.~Grynberg and C.~Robilliard, Phys. Rep.
         \textbf{355}, 335 (2001).
         \bibitem{steck01} D.A. ~Steck, W.H.~Oskay and M.G.~Raizen, Science
         \textbf{293}, 274 (2001) and references therein.
         \bibitem{hensinger01} W.K. Hensinger, H. H{\"a}ffner, A. Browaeys, N.R.
         Heckenberg, K. Helmerson, C. McKenzie, G.J. Milburn, W.D. Phillips,
   S.L.
         Rolston, H.~Rubinstein-Dunlop, and B.~Upcroft, Nature \textbf{ {412}},
   52
         (2001) and references therein.
         \bibitem{deutsch98} See, e.g., I.H.~Deutsch and P.S.~Jessen,
         Phys. Rev. A \textbf{57}, 1972 (1998), and D.~Jaksch, H.-J. Briegel,
         J.I.~Cirac, C.W.~Gardiner,
         and P.~Zoller, Phys. Rev. Lett. \textbf{ {82}}, 1975 (1998).

       \bibitem{kurizki02} G. Kurizki, S. Giovanazzi, D. O'Dell, A. I.
         Arte\-miev\emph{New Regimes in Cold Gases Via Laser-Induced
           Long-Range Interactions}, in ``Dynamics and Thermodynamics
         of Systems with Long Range Interactions'', T. Dauxois, S.
         Ruffo, E. Arimondo, M. Wilkens Eds.,  Lecture Notes in Physics
         Vol. 602, Springer (2002), (in this volume)

         \bibitem{hemmerich93}A.~Hemmerich and T.W.~H{\"a}nsch, Phys. Rev.
         Lett. \textbf{ {70}}, 410 (1993).
         \bibitem{grynberg93}G.~Grynberg, B.~Lounis, P.~Verkerk,
         J.Y.~Courtois, and C.~Salomon, Phys. Rev. Lett. \textbf{70}, 2249
         (1993).
         \bibitem{kittel} see, e.g., C.~Kittel, \emph{Introduction to
            Solid State Physics} (Wiley, New York, 1996).
            \bibitem{verkerk92}P.~Verkerk, B.~Lounis, C.~Salomon,
         C.~Cohen-Tannoudji, J.Y.~Courtois, and G.~Grynberg, Phys. Rev.
         Lett. \textbf{68}, 3861 (1992).
         \bibitem{rudy97}P.~Rudy, R.~Ejnisman, and N.P.~Bigelow, Phys. Rev.
         Lett. \textbf{78}, 4906 (1997).
         \bibitem{raithel98}G.~Raithel, W.D.~Phillips, and S.L.~Rolston,
         Phys. Rev. Lett. \textbf{81}, 3615 (1998).
         \bibitem{buchkremer00}F.B.J.~Buchkremer, R.~Bumke, H.~Levsen,
         G.~Birkl, and W.~Ertmer, Phys. Rev. Lett. \textbf{85}, 3121 (2000).
         \bibitem{triche96}C.~Trich{\'e}, L.~Guidoni, P.~Verkerk, and
         G.~Grynberg, in \emph{OSA TOPS on ultra-cold atoms and BEC}, edited by
         K.~Burnett, vol. 7, p. 82,(OSA, Washington D.C., 1996).
         \bibitem{morsch00}O.~Morsch, P.H.~Jones, and D.R.~Meacher, Phys.
         Rev. A \textbf{ {61}}, 023410 (2000).
         \bibitem{weidemuller95}M.~Weidem{\"u}ller, A.~Hemmerich,
         A.~G{\"o}rlitz, T.~Esslinger, and T.W.~H{\"a}nsch, Phys. Rev. Lett.
         \textbf{75}, 4583 (1995).
         \bibitem{birkl95}G.~Birkl, M.~Gatzke, I.H.~Deutsch, S.L.~Rolston,
         and W.D.~Phillips, Phys. Rev. Lett. \textbf{75}, 2823 (1995).
         \bibitem{gorlitz97}A.~G{\"o}rlitz, M.~Weidem{\"u}ller, T.W.~H{\"a}nsch,
         and A.~Hemmerich, Phys. Rev. Lett. \textbf{78}, 2096 (1997).
         \bibitem{meacher95}D.R.~Meacher, S.~Guibal, C.~Mennerat,
         J.Y.~Courtois, K.I.~Petsas, and G.~Grynberg, Phys. Rev. Lett.
      \textbf{74}, 1958 (1995).
         \bibitem{courtois96}J.Y.~Courtois, S.~Guibal, D.R.~Meacher,
         P.~Verkerk, and G.~Grynberg, Phys. Rev. Lett. \textbf{77}, 40 (1996).
       \bibitem{sanchez02}L. Sanchez-Palencia, F.-R. Carminati, M. Schiavoni,
       F. Renzoni, and G. Grynberg , Phys. Rev. Lett. \textbf{88}, 133903
   (2002).
         \bibitem{guidoni99} L. Guidoni, B. D{\'e}pret, A. di Stefano, and P.
         Verkerk, Phys. Rev. A \textbf{60}, R4233 (1999).
         \bibitem{hemmerich95}A.~Hemmerich, M.~Weidem{\"u}ller, T.~Esslinger,
         C.~Zimmermann, and T.W.~H{\"a}nsch, Phys. Rev. Lett. \textbf{75}, 37
         (1995).
         \bibitem{hamann98}S.E.~Hamann, D.L.~Haycock, G.~Klose, P.H.~Pax,
         I.H.~Deutsch, and P.S.~Jessen, Phys. Rev. Lett. \textbf{80}, 4149
         (1998).
         \bibitem{perrin98}H.~Perrin, A.~Kuhn, I.~Bouchoule, and
         C.~Salomon, Europhys. Lett. \textbf{42}, 395 (1998).
         \bibitem{vuletic98}V.~Vuletic, C.~Chin, A.J.~Kerman, and S.~Chu,
         Phys. Rev. Lett. \textbf{81}, 5768 (1998).
         \bibitem{madison99}K.W.~Madison, M.C.~Fischer, and M.G.~Raizen,
         Phys. Rev. A \textbf{60}, R1767 (1999).
         \bibitem{wilkinson97}S.R.~Wilkinson, C.F.~Bharucha, M.C.~Fischer,
         K.W.~Madison, P.R.~Morrow, Qian Niu, B.~Sundaram, and M.G.~Raizen,
         Nature \textbf{ {387}}, 575 (1997).
         \bibitem{fischer01}M.C.~Fischer, B.~Guttierez-Medina, and
         M.G.~Raizen, Phys. Rev. Lett. \textbf{ {87}}, 040402 (2001).
         \bibitem{mennerat99}C.~Mennerat-Robilliard, D.~Lucas, S.~Guibal,
         J.~Tabosa, C.~Jurczak, J.Y.~Courtois, and G.~Grynberg, Phys. Rev.
         Lett. \textbf{ {82}}, 851 (1999).
         \bibitem{bendahan96} M.~Ben-Dahan, E. Peik, J. Reichel, Y. Castin,
         and C. Salomon, Phys. Rev. Lett \textbf{ {76}},
         4508 (1996);
         \bibitem{peik} E.~Peik, M.~Ben~Dahan, I.~Bouchoule, Y.~Castin, and
         C.~Salomon, Phys. Rev.
            A \textbf{ {55}}, 2989 (1997).
      \bibitem{wilkinson96} S.R.~Wilkinson, C.F.~Bharucha, K.W.~Madison,
         Qian~Niu, and M.G.~Raizen, Phys. Rev. Lett. \textbf{ {76}}, 4512
   (1996).
      \bibitem{niu96} Qian~Niu, Xian-Geng~Zhao, G.A.~Georgakis, and
         M.G.~Raizen, Phys. Rev. Lett. \textbf{ {76}}, 4504 (1996).
      \bibitem{raizen97} M.~Raizen,C.~Salomon, and Qian Niu, Phys. Today
   \textbf{
         50} (7), 30 (1997).

         \bibitem{bharucha97} C.F.~Bharucha, K.W.~Madison, P.R.~Morrow,
         S.R.~Wilkinson, B.~Sundaram, and M.G.~Raizen, Phys. Rev. A \textbf{
         {55}}, R857 (1997).

         \bibitem{inguscio} See reviews by W. Ketterle, D.S. Durfee and D.M.
         Stamper-Kurn, and by E.~Cornell {\it et al.} in \emph{Bose-Einstein
         condensation in atomic gases}, edited by M.~Inguscio, S.~Stringari
         and C.~Wieman (IOS Press, Amsterdam) (1999).
         \bibitem{sorensen98} K.~Berg-S{\o}rensen and K.~M{\o}lmer, Phys.
         Rev. A \textbf{ 58}, 1480 (1998).
         \bibitem{jaksch98} D.~Jaksch, C.~Bruder, J.I.~Cirac, C.W.~Gardiner,
         and P.~Zoller, Phys. Rev. Lett. \textbf{ 81}, 3108 (1998).
         \bibitem{kozuma99} M.~Kozuma, L.~Deng, E.W.~Hagley, J.~Wen,
         R.~Lutwak, K.~Helmerson, S.L.~Rolston, and W.D.~Phillips, Phys.
         Rev. Lett. \textbf{ 82}, 871 (1999).
         \bibitem{ovchinnikov99} Yu.B.~Ovchinnikov, J.H.~M{\"u}ller,
         M.R.~Doery, E.J.D.~Vredenbregt, K.~Helmerson, S.L.~Rolston, and
         W.D.~Phillips, Phys. Rev. Lett. \textbf{ 83}, 284 (1999).
         \bibitem{hagley99} E.W.~Hagley, L. Deng, M. Kozuma, M. Trippenbach, Y.
   B.
         Band,
         M. Edwards, M. Doery, P. S. Julienne, K. Helmerson, S. L. Rolston, and
   W.
      D.
         Phillips,
         Phys. Rev. Lett. \textbf{ 83}, 3112 (1999).
         \bibitem{stenger99} J.~Stenger, S. Inouye,
         A. P. Chikkatur, D. M. Stamper-Kurn, D. E. Pritchard, and W. Ketterle,
         Phys. Rev. Lett. \textbf{82}, 4569 (1999); and erratum: Phys. Rev.
   Lett.
         \textbf{84}, 2283 (2000).
         \bibitem{simsarian00} J.E.~Simsarian, J.~Denschlag,
         M.~Edwards, C.W.~Clark, L.~Deng, E.W.~Hagley, K.~Helmerson,
   S.L.~Rolston,
         and
         W.D.~Phillips, Phys. Rev. Lett. \textbf{ 85}, 2040 (2000).
         \bibitem{vogels02} J. M. Vogels, K. Xu, C. Raman, J. R. Abo-Shaeer,
   and
         W. Ketterle, Phys. Rev. Lett. 88, 060402 (2002)
         \bibitem{steinhauer02} J. Steinhauer, R. Ozeri, N. Katz, and N.
         Davidson, Phys. Rev. Lett. 88, 120407 (2002).
         \bibitem{ozeri02} R.~Ozeri, J.~Steinhauer, N.~Katz, and
            N.~Davidson, Phys. Rev. Lett. 88, 220401 (2002).
         \bibitem{anderson98} B.P.~Anderson and M.~Kasevich, Science \textbf{
         282}, 1686 (1998).
         \bibitem{orzel01} C. Orzel, A.K.~Tuchman, M.L.~Fenselau,
         M.~Yasuda, and M.A.~Kasevich, Science \textbf{ 231}, 2386
            (2001).
         \bibitem{greiner02} M.~Greiner, O.~Mandel, T.~Esslinger,
            T.W.~H{\"a}nsch, and I.~Bloch, Nature \textbf{ 415}, 6867 (2002).
         \bibitem{burger01} S.~Burger, F.~Cataliotti, C.~Fort, F.~Minardi,
         M.~Inguscio, M.L.~Chiofalo, and
            M.P.~Tosi, Phys. Rev. Lett. \textbf{ 86}, 4447 (2001).
         \bibitem{cataliotti01} F.S.~Cataliotti, S.~Burger, C.~Fort,
            P.~Maddaloni, F.~Minardi, A.~Trombettoni, A.~Smerzi, and
            M.~Inguscio, Science \textbf{ 293}, 843 (2001).
         \bibitem{morsch01} O.~Morsch, J.H.~M{\"u}ller, M.~Cristiani,
            D.~Ciampini, and E.~Arimondo, Phys. Rev. Lett. \textbf{ 87}, 140402
            (2001).
         \bibitem{cristiani02} M. Cristiani, O. Morsch, J. H. MTller, D.
   Ciampini,
         and E. Arimondo, Phys. Rev. A \textbf{ 65}, 063612 (2002).
         \bibitem{pedri01} P.~Pedri, L.~Pitaevskii, S.~Stringari,
            C.~Fort, S.~Burger, F.S.~Cataliotti, P.~Maddaloni, F.~Minardi,
            and M.~Inguscio, Phys. Rev. Lett. \textbf{ 87}, 220401 (2001).
         \bibitem{morsch02} O.~Morsch, M. Cristiani, J.H.~M{\"u}ller,
            D.~Ciampini, and E.~Arimondo,cond-mat 0204528.
         \bibitem{denschlag02} J. Hecker-Denschlag, J. E. Simsarian, H.
   H{\"a}ffner,
         C. McKenzie, A. Browaeys, D. Cho, K. Helmerson,
         S. L. Rolston, and W. D. Phillips, J. Phys. B: At. Mol. Opt. Phys in
      press (2002).
         \bibitem{burger01b} S. Burger, F. S. Cataliotti, C. Fort, P.
   Maddaloni,
         F. Minardi and M. Inguscio, Europhys. Lett. \textbf{ 57} 1 (2002).
         \bibitem{greiner01} M.~Greiner, I.~Bloch, O.~Mandel, T.W.~H{\"a}nsch,
         and T. Esslinger, Phys. Rev. Lett. \textbf{ 87}, 160405 (2001).

         \bibitem{burger99} S.~Burger, K.~Bongs, S.~Dettmer, W.~Ertmer,
            K.~Sengstock, A.~Sanpera, G.V.~Shlyapnikov, and M.~Lewenstein,
   Phys.
            Rev. Lett. \textbf{ 83}, 5198 (1999).
         \bibitem{deng99} L.~Deng, E.W.~Hagley, J.~Wen, M.~Trippenbach,
   Y.~Band,
            P.S.~Julienne, J.E.~Simsarian, K.~Helmerson, S.L.~Rolston, and
            W.D.~Phillips, Nature \textbf{ 398}, 6724 (1999).
         \bibitem{dalfovo99} F.~Dalfovo, S.~Giorgini, L.P.~Pitaevskii,
            and S.~Stringari, Rev. Mod. Phys. \textbf{ 71}, 463 (1999).
         \bibitem{choi99} D.I.~Choi and Qian Niu, Phys. Rev. Lett. \textbf{
         82}, 2022 (1999).
          \bibitem{brunello00} A.~Brunello, F.~Dalfovo, L.~Pitaevskii, and
         S.~Stringari, Phys. Rev. Lett.
         \textbf{ 85}, 4422 (2000).
         \bibitem{chiofalo01} M.L.~Chiofalo, S. Succi, and M.P. Tosi, Phys.
   Rev.
         A \textbf{ 63}, 063613 (2001).
         \bibitem{kramer02} M.Kr{\"a}mer, L. Pitaevskii, and S. Stringari, Phys.
         Rev. Lett. \textbf{88}, 180404 (2002).
         \bibitem{bronski01} J.C.~Bronski, L.D.~Carr, B.~Deconink, and J.N.
         Kutz, Phys. Rev. lett. \textbf{86}, 1402 (2001).
         \bibitem{wu02} B. Wu, R.B.~Diener, and Qian Niu, Phys. Rev. A
   \textbf{65},
         025601 (2002).
         \bibitem{diakonov01} D.Diakonov, L.M.~Jensen, C.J.~Pethick, and H.
         Smith, e-print: cond-mat/0111303.
         \bibitem{wu00} B. Wu and Qian Niu, Phys. Rev. A \textbf{61}, 023401
      (2000).
         \bibitem{zobay00} O.Zobay and B.M. Garraway, Phys. Rev. \textbf{61}
         033603 (2000).
         \bibitem{trombettoni01} A.~Trombettoni and A. Smerzi, Phys. Rev. Lett.
         \textbf{86}, 2353 (2001).
         \bibitem{wu01} B. Wu and Qian Niu, Phys. Rev. A \textbf{64}, 061603
      (2001).
         \bibitem{pitaevskii01} L. Pitaevskii, and S. Stringari, Phys.
         Rev. Lett. \textbf{88}, 180402 (2001).
         \bibitem{cuccoli01} A. Cuccoli, A. Fubini, V. Tognetti, and R. Vaia,
      Phys.
         Rev. A \textbf{64}, 061601 (2001).
         \bibitem{vanoosten} D. van Oosten, P. van der Straten, and H.T.C.
         Stoof, Phys. Rev. A \textbf{63} 053601 (2001).
         \bibitem{burnett02} K.Burnett, M. Edwards, C.W. Clark, and M. Shotter,
         J. Phys. B: At. Mol. Opt. Phys. \textbf{35}, 1671 (2002).

         \bibitem{abdullaev01} F. Kh. Abdullaev, B. B. Baizakov, S. A.
   Darmanyan,
      V.
         V. Konotop,
         and M. Salerno, Phys. Rev. A \textbf{64}, 043606 (2001).
         \bibitem{konotop01} V.V.~Konotop and M.Salerno, Phys. Rev. A
   \textbf{65},
         021602(R) (2002).
         \bibitem{tsukada02} N. Tuskada, Phys. Rev. A \textbf{65}, 063608
   (2002).
         \bibitem{kalinski} M.~Kalinski, G. Metakis, and W.P. Schleich, private
         communication.




         \end{thebibliography}
    \end{document}